\documentclass{elsart}

 \usepackage{graphicx}

\usepackage{amssymb}

\def\beq{\begin{equation}}
\def\eeq{\end{equation}}
\def\bea{\begin{eqnarray}}
\def\eea{\end{eqnarray}}

\newcommand*{\eqref}[1]{Eq.~(\ref{eq:#1})}

\newcommand*{\figref}[1]{Fig.~\ref{fig:#1}}
\newcommand*{\figlab}[1]{\label{fig:#1}}

\def\VYP#1#2#3{{\bf #1}, #3 (#2)}  

\def\PLB#1#2#3{Phys.~Lett.~B~\VYP{#1}{#2}{#3}}

\def\PRD#1#2#3{Phys.~Rev.~D~\VYP{#1}{#2}{#3}}
\def\PRL#1#2#3{Phys.~Rev.~Lett.~\VYP{#1}{#2}{#3}}

\newcommand{\Omit}[1]{}

\begin{document}

\begin{frontmatter}



\title{}


\title{Status report of the NuMoon experiment.}

\author[KVI]{O. Scholten}
\ead{scholten@kvi.nl}
\author[Nym]{S. Buitink}
\author[ASML]{J. Bacelar}
\author[CSIRO]{R. Braun}
\author[Kapt,ASTRON]{A.G. de Bruyn}
\author[Nym]{H. Falcke}
\author[KVI]{K. Singh}
\author[Man]{B. Stappers}
\author[UvA,ASTRON]{R.G. Strom}
\author[KVI]{R. al Yahyaoui}

\address[KVI]{Kernfysisch Versneller Instituut, University of Groningen, 9747 AA,
Groningen, The Netherlands}

\address[Nym]{Department of Astrophysics, IMAPP, Radboud University, 6500 GL Nijmegen, The Netherlands}

\address[ASML]{ASML Netherlands BV, P.O.Box 324, 5500 AH Veldhoven, The
Netherlands}

\address[CSIRO]{CSIRO-ATNF, P.O.Box 76, Epping NSW 1710, Australia}

\address[Kapt]{Kapteyn Institute, University of Groningen, 9747 AA, Groningen, The Netherlands}

\address[ASTRON]{ASTRON, 7990 AA Dwingeloo, The Netherlands}

\address[Man]{School of Physics \& Astronomy, Alan Turing Building, Univ. of Manchester, Manchester, M13 9PL}

\address[UvA]{Astronomical Institute `A. Pannekoek', University of Amsterdam, 1098
SJ, The Netherlands}

\begin{abstract}
We show that at wavelengths comparable to the length of the shower produced by an
Ultra-High Energy cosmic ray or neutrino, radio signals are an extremely
efficient way to detect these particles. First results are presented of an
analysis of 20 hours of observation data for NuMoon project using the Westerbork
Synthesis Radio Telescope to search for short radio pulses from the Moon. A limit
on the neutrino flux is set that is a factor four better than the current one
(based on FORTE).
\end{abstract}

\end{frontmatter}

\section{}
\label{}

\section{Introduction}

An efficient method to determine the fluxes of Ultra High Energy (UHE) particles
is through the production of coherent radio waves~\cite{Ask62} when an UHE
particle hits the moon. At sufficiently high energy the pulses are detectable at
Earth with radio telescopes, an idea first proposed by Dagkesamanskii and
Zheleznyk~\cite{Dag89}. Several experiments have since been
performed~\cite{Han96,Gor04}. These experiments have looked for radiation near
the frequency where the intensity is expected to reach its maximum. Since the
typical lateral size of a shower is of the order of 10~cm the peak frequency is
of the order of 3~GHz.

We propose~\cite{Sch06} to look for the radio waves at considerably lower
frequencies where the wavelength of the radiation is comparable in magnitude to
the typical longitudinal size of showers. As discussed in Section {\bf 3} the
lower intensity of the emitted radiation is compensated by an increase in
detection efficiency due to the near isotropic emission of coherent radiation,
resulting in an increase of sensitivity by several orders of magnitude. In
Section {\bf 4} we discuss the on-going observations made for the NuMoon project
with the Westerbork Synthesis Radio-Telescope array (WSRT) and the neutrino-flux
limit which has been determined from 20 hours observation time.

\section{Radio Emission}

The intensity of radio emission from a hadronic shower with energy $E_s$ can be
parameterized as given in Ref.~\cite{Sch06}. The angle at which the intensity of
the radiation reaches a maximum, the \v{C}erenkov angle, is related to the index
of refraction ($n$) of the medium, $\cos{\theta_c}=1/n$. Since the angular spread
of the intensity is crucial for our considerations, we have derived analytic
formulas for a ``block'' and a ``sine'' shower profile~\cite{Sch06,Leh04}, curves
labeled respectively 'b' and 's' in \figref{ang-spread}. For a shower of
$10^{20}$~eV It is seen that at 2.2 GHz the simple exponential
form~\cite{Gor04,Alv01,Zas92} approximates well the analytic forms.  At 100 MHz
(right hand panel of \figref{ang-spread}) the analytic expressions all give
similar results differing from the phenomenological exponential parameterization.
The reason for this difference lies mainly in the pre-factor $\sin^2{\theta}$
which accounts for the radiation being polarized parallel to the shower, not
allowing for emission at 0$^\circ$ and 180$^\circ$.
\begin{figure}
    \includegraphics[height=7.9cm,bb=50 147 515 650,clip]{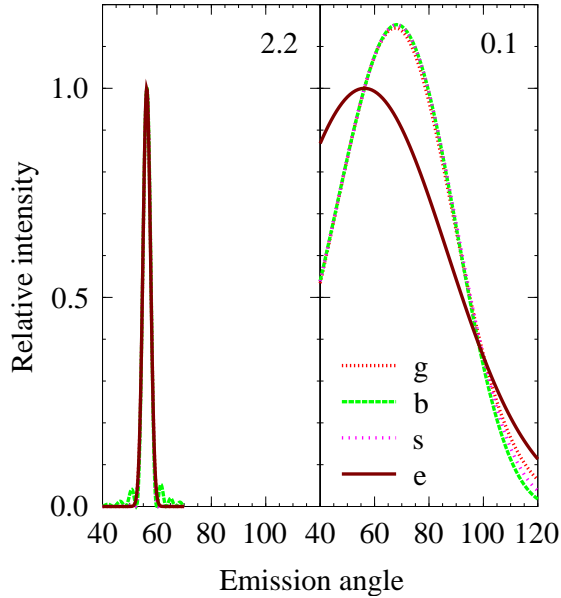}
\hspace{2pc} \begin{minipage}[b]{13pc}
\caption[fig9]{The angular spread around the \v{C}erenkov angle for different
shower-profile functions (see text) are compared to the parametrization used in
this work. The left (right) hand displays the results for 2.2 GHz (100MHz)
respectively. }
  \figlab{ang-spread}
\end{minipage}
\end{figure}

Cosmic-ray-induced showers occur effectively at the lunar surface. For
neutrino-induced showers an energy-dependent mean free path has been used. The
attenuation length for the radiated power in the regolith is set at $\lambda_r=
(9/\nu$[GHz])~m. A crucial point in the simulation is the refraction of radio
waves at the lunar surface~\cite{Zas92}. The emitted radiation at high
frequencies where the \v{C}erenkov cone is rather narrow will be severely
diminished  due to internal reflection at the surface. The major advantage of
going to lower frequencies is that the angular spreading increases, allowing the
radiation to penetrate the lunar surface. With decreasing frequency the intensity
of the emitted radiation decreases, however, the intensity increases with
increasing particle energy. The net effect is that at sufficiently-high shower
energies the aforementioned effect of increased spreading is far more important,
resulting in a strong increase in the detection probability, see \figref{Flux}.
An additional advantage of using lower frequencies is that the sensitivity of the
model simulations to large- or small-scale surface roughness is diminished  since
the angular spread is already large. This is in contrast to high frequencies
where most of the radiation is internally reflected when surface roughness is
ignored.
The thickness of the regolith is about 10--100~m and known to vary over the lunar
surface. There is a (probably smooth) transition to solid rock, for which the
density is about twice that of the regolith. The crust appears to be homogeneous
in composition up to depths of about 20~km where there is a seismic
discontinuity~\cite{Wie01}. Pure rock and regolith have been simulated and found
to give very similar detection limits for low frequencies~\cite{Sch06}.

\begin{figure}
    \includegraphics[height=7.9cm,bb=36 137 515 672,clip]{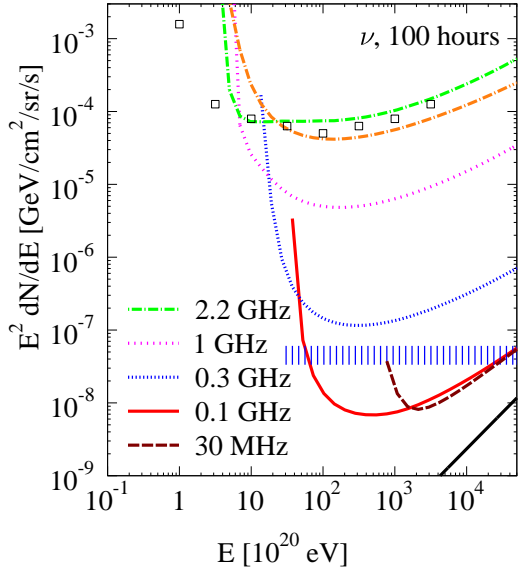}
\hspace{2pc} \begin{minipage}[b]{13pc}
\caption[fig2]{Flux limits (assuming a null observation) for
neutrinos  as can be determined in a 100 hour observation.
In the curves for $\nu=30$~MHz a ten fold higher detection threshold is
used, corresponding to the higher sky temperature at this frequency.
The thick black line corresponds to the best possible limit
(vanishing detection threshold). The open squares for the neutrino flux are the
limits determined from the GLUE experiment~\cite{Gor04}.}
  \figlab{Flux}
\end{minipage}
\end{figure}

\section{WSRT observations}

The Westerbork Radio Synthesis Telescope (WSRT) is an array telescope consisting
of 14 parabolic telescopes of 25~m on a 2.7~km east-west line. The NuMoon
experiment uses the Low Frequency Front Ends (LFFEs) which cover the frequency
range 115--180 MHz and record full polarization data. For our observations we use
the Pulsar Machine II backend~\cite{kss08}, which can record 8 Nyquist sampled
bands with a bandwidth of 20 MHz each (40 MHz sampling frequency). We use two
beams of 4 bands each, centered around 123, 137, 151, and 165 MHz. The two beams
are aimed at different sides of the Moon, each covering about one third of the
lunar surface. A real lunar Cherenkov pulse should only be visible in only one of
the two beams. Because of overlap the total bandwidth per beam is 65 MHz.

\begin{figure}
 \includegraphics[width=.4\linewidth,angle=270]{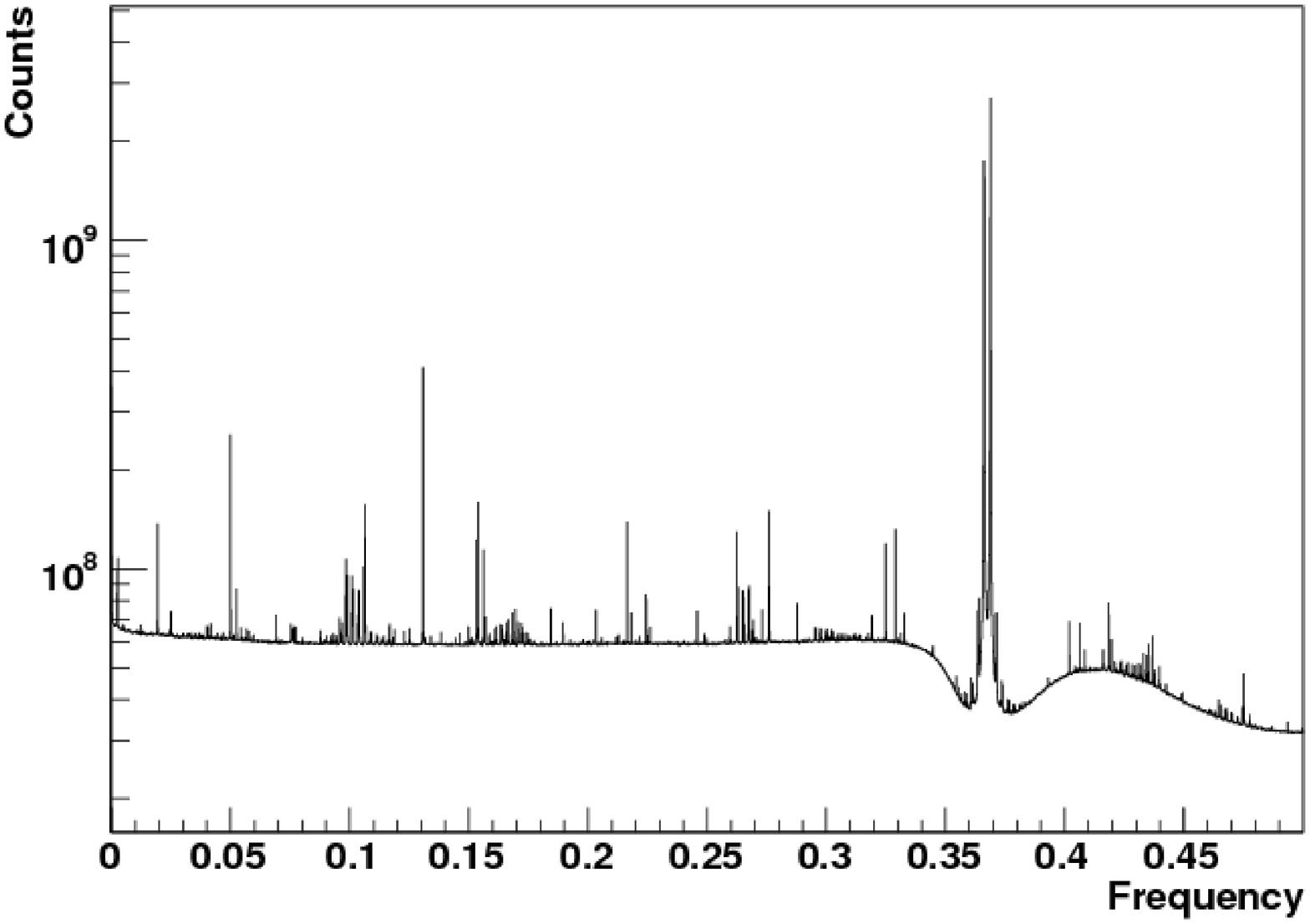}
 \includegraphics[width=.4\linewidth, angle=270]{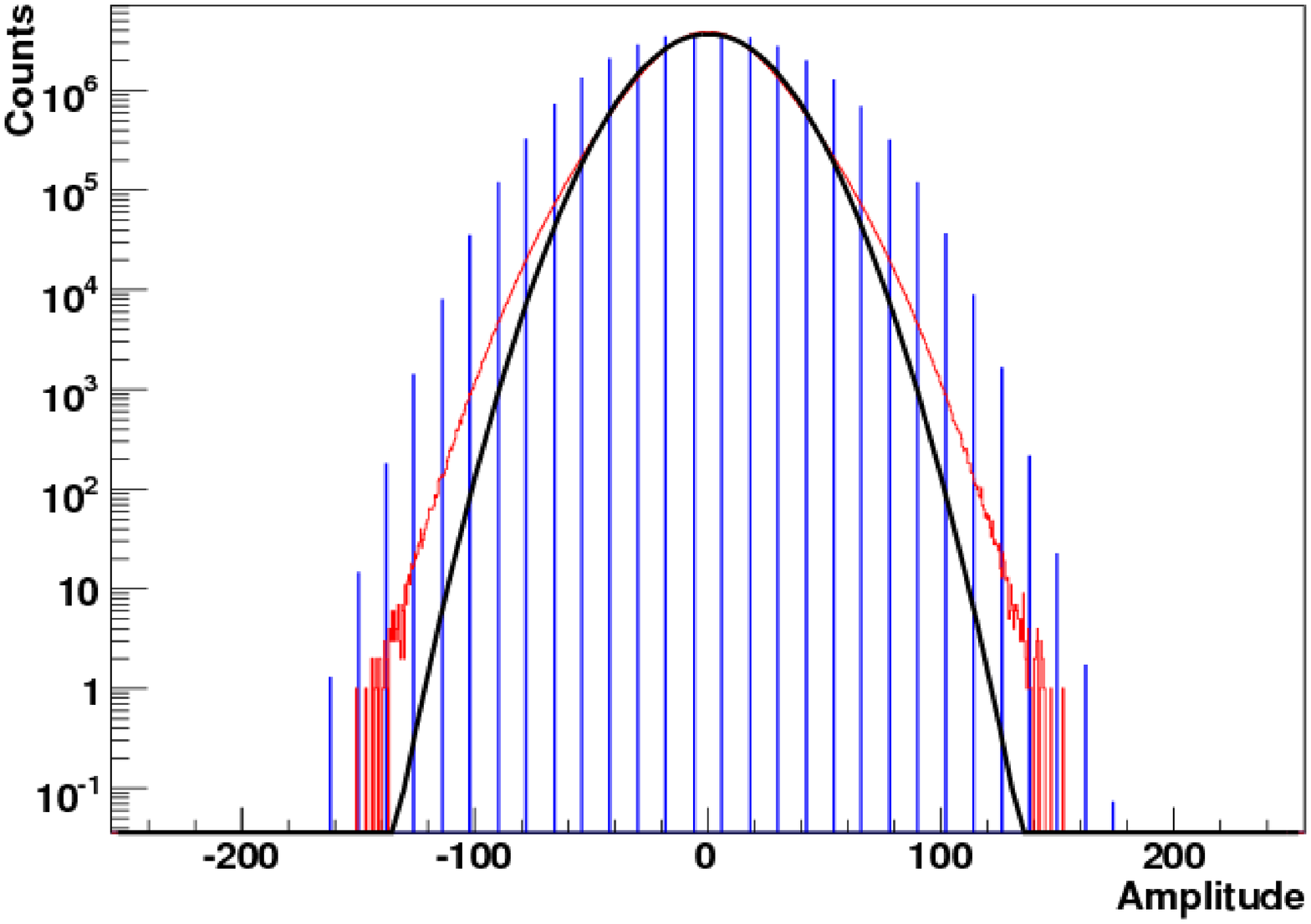}
\caption{Left: Typical frequency spectrum of 10 seconds of data before RFI reduction.
The x-axis shows arbitrary units. Right: Typical results of RFI
reduction. The number of counts per amplitude is plotted for the raw data (blue) and the data after RFI reduction (red).
The black line is a Gaussian fit to the data after RFI reduction.}
\label{fig:analysis}
\end{figure}

In the first pass a rough search for pulses is performed to retain about 1\% of
the data,  stored permanently. The procedure involves the following
steps~\cite{Bui08}:
\begin{enumerate}
\item  The raw data is read in blocks of 20\,000 time bins each and Fourier
    transformed to excise narrow-band radio-frequency interference (RFI). The
    left panel in Fig. \ref{fig:analysis} shows the frequency spectrum for
    band centered around 165 MHz of 10 seconds of data before RFI removal.
    The right hand side shows the effects of RFI removal.
\item The data is transformed back to the time domain after a de-dispersion
    is performed, based on the slanted total electron content (STEC) of the
    ionosphere. This corrects for the dispersion of signals coming from the
    Moon.
\item The pulses of interest have a width smaller than the bin size (25~ns)
    and should thus be sharp in the de-dispersed data. However a
    Nyquist-sampled, bandwidth limited, pulse can be three time-samples long.
    In addition a mismatch in the STEC value used for de-dispersion will
    further broaden the pulse. We define the value $P5$ as the power
    integrated over 5 time bins and 2 polarizations, normalized to the
    average value of this integration, $P5=P5_x +P5_y$ with $
    P5_i=\sum_{\mathrm{5\ bins}} P_i/{\bigg<\sum_{\mathrm{5\ bins}} P_i
    \bigg>} $ where the
averaging is done per block.
\item A peak search is carried out, where we use a trigger condition of a
    $P5$ value of 2.5 in all four frequency bands. A time difference between
    the peaks in the different frequency bands is allowed to account for a
    possible error in the STEC value used in de-dispersion. For each trigger
 the data blocks in which the pulses are found are
    stored for postprocessing.
\end{enumerate}

In the second pass, further cuts are applied to the data. Pulses of a width
exceeding 8 bins are rejected, as well as pulses that co\"{i}ncide with a regular
timer pulse.

\section{Results}
\begin{figure}
\centerline{\includegraphics[width=.85\linewidth]{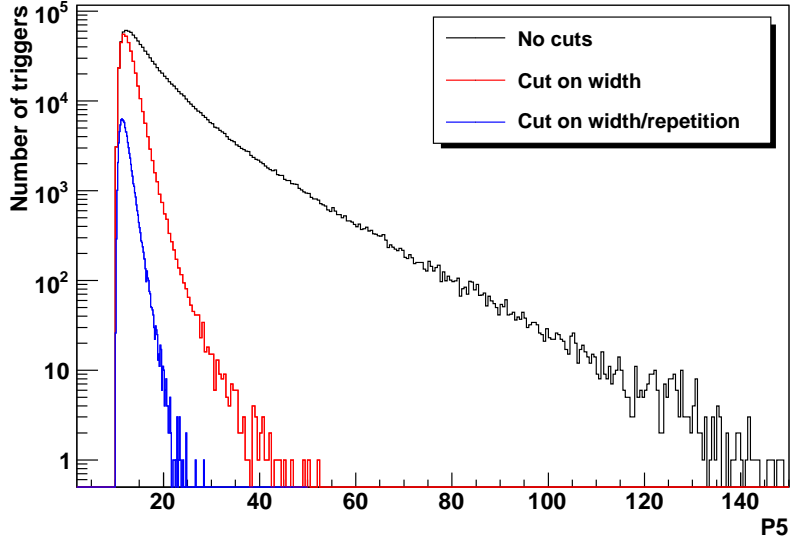}}
\caption{Number of triggers plotted against the P5 value for different cuts.
\label{fig:cuts}}
\end{figure}
Presently, 10 hours and 40 minutes of single beam data have been accumulated.
Fig. \ref{fig:cuts} shows the distribution of $P5$ values with no cuts applied
(black line), a cut on pulse width (red line) and a cut on the timing signal
(green line). For pure Gaussian noise the number of expected triggers is of the
order of unity at $P5\approx 14$.

The highest value surviving this cut is $P5=25$. For the WSRT the system noise at
low frequencies is $F_{noise}=600$~Jy per polarization channel, so $P5=1$
corresponds to 6000~Jy, giving a detection threshold of $\sim 38$~kJy.
\figref{fig:limit} shows the limit on the neutrino flux that is obtained with the
current 20 hours of data based on this detection threshold. Also shown is the
limit that can be achieved with a 100 hour observation period. The current limits
in the UHE region are established by ANITA \cite{Bar06} and FORTE \cite{Leh04}.
Two model predictions are plotted: the Waxman-Bahcall limit \cite{Bah01} and a
top-down model \cite{Pro96} for exotic particles of mass $M_X=10^{24}$~eV.

\begin{figure}
    \includegraphics[width=.5\linewidth,bb=55 165 540 680,clip]{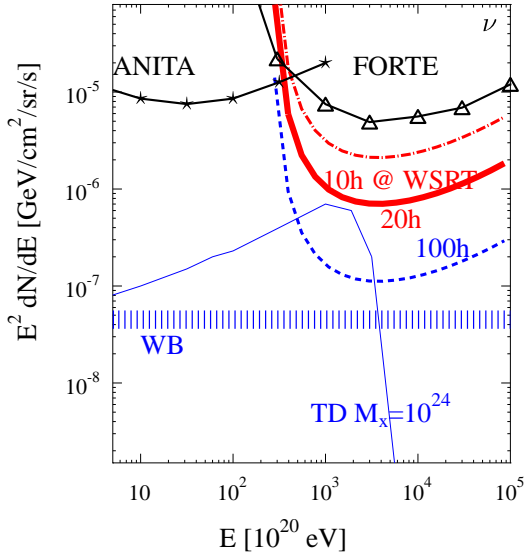}
\hspace{2pc} \begin{minipage}[b]{13pc}
\caption[fig5]{Neutrino flux limit currently established with 20 hours of WSRT data and the limit that will be achieved
with 100 hours. Limits from ANITA and FORTE
are included in the plot as well as the Waxman-Bahcall flux and a TD model prediction.}
  \figlab{fig:limit}
\end{minipage}
\end{figure}

\section{Discussion}
We observe at a frequency window that offers an optimal sensitivity to lunar
pulses. Because of the large spread in emission angle, we expect no systematic
effect from surface irregularities. The detection efficiency is also largely
independent from details in the structure of the (sub-)regolith \cite{Sch06}.
With about 100 hours of observation data we will be able to put a limit on the
UHE neutrino flux which is about an order of magnitude lower than the current
FORTE limit (assuming no detections). This limit would rule out a subset of TD
models (see \figref{fig:limit}).
\begin{figure}
\includegraphics[width=.5\linewidth,bb=41 136 512 656,clip]{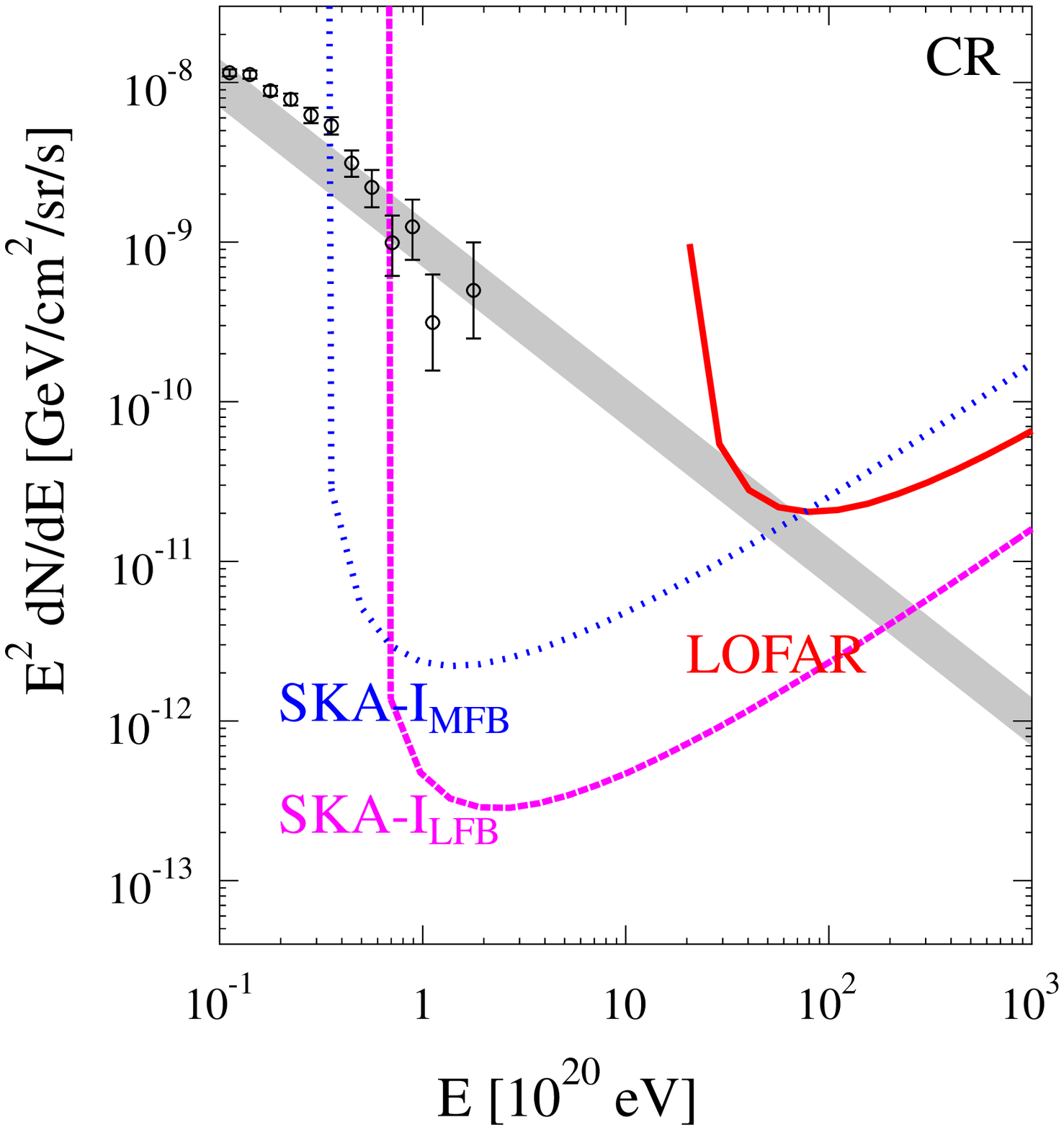}
\includegraphics[width=.5\linewidth,bb=41 136 512 656,clip]{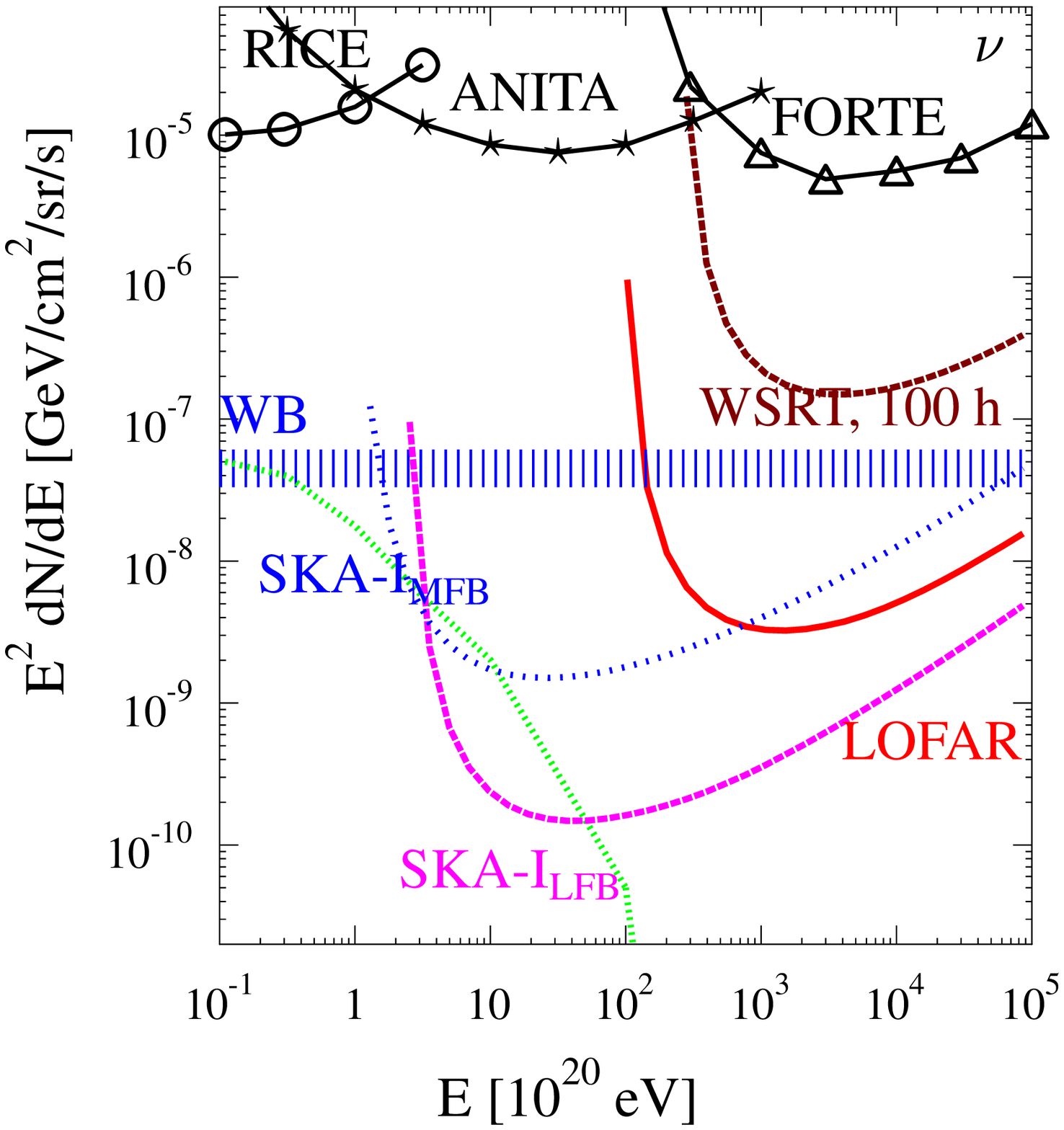}
\caption{Limits on UHE neutrino flux (left) and cosmic ray flux (right) that can be established with LOFAR and
SKA.}
\figlab{lofar}
\end{figure}

The next phase in the NuMoon project will be to use LOFAR, the Low Frequency
Array, that is under construction in the Netherlands. LOFAR is a network of low
frequency omni-directional radio antennas communicating over a fiber optics
network. Half of the stations are located inside the 2~km$\times$2~km core with a
total collecting area of $\sim$0.05~km$^2$. Multiple beams can be formed to cover
the surface of the Moon, resulting in a sensitivity that is about 25 times better
than the WSRT.  \figref{lofar} shows the sensitivity that will be achieved with
30 days of observation time with LOFAR for UHE cosmic rays (left panel) and
neutrinos (right panel).

Other lunar Cherenkov observations will be carried out at the Australia Telescope
Compact Array (ATCA) \cite{james} consisting of six 22~m dishes. The array is
currently undergoing an upgrade to be able to measure with a bandwidth of 2 GHz.
The upgrade is projected to be finished in 2009.

Eventually, the best sensitivity will be achieved with the Square Kilometer Array
\cite{ska} (SKA), planned to be completed in 2020. The Australian SKA Pathfinder
(ASKAP) is expected to be operational around 2011.  In \figref{lofar} the
expected sensitivity of SKA is plotted for observations in the low frequency band
(70--200~MHz) and the middle frequency band (200--300~MHz).

 This work was performed as part of the research programs of the Stichting
voor Fundamenteel Onderzoek der Materie (FOM) and of ASTRON, both with financial
support from the Nederlandse Organisatie voor Wetenschappelijk Onderzoek (NWO).

\end{document}